\RequirePackage{fix-cm}
\documentclass[smallextended]{svjour3}       % onecolumn (second format)
\smartqed  % flush right qed marks, e.g. at end of proof
\usepackage{cite}
\usepackage{pifont}%

\usepackage{inconsolata}
\usepackage{graphicx}
\usepackage{wrapfig}
\usepackage{url}
\usepackage{adjustbox}
\usepackage{colortbl}
\usepackage[usenames,dvipsnames]{xcolor}
\usepackage{tikz}
\definecolor{alizarin}{rgb}{0.82, 0.1, 0.26}

\newcommand*\circled[1]{\tikz[baseline=(char.base)]{
            \node[minimum width=1pt, shape=circle,fill=black,inner sep=1pt] (char) {{\footnotesize \textcolor{white}{#1}}};}}

\newcommand{\one}{\circled{a}}
\newcommand{\two}{\circled{b}}
\newcommand{\three}{\circled{c}} 
\newcommand{\four}{\circled{d}}
\newcommand{\five}{\circled{e}}
\newcommand{\six}{\circled{f}}
\newcommand{\seven}{\circled{g}}
\newcommand{\eight}{\circled{h}}
\newcommand{\nine}{\circled{i}}
\newcommand{\ten}{\circled{j}}
\newcommand{\eleven}{\circled{k}}
\newcommand{\twelve}{\circled{l}}
\newcommand{\thirteen}{\circled{m}}

\definecolor{CustomDarkRed}{RGB}{175, 0, 0} 
\newcommand{\RED}[1]{{\color{CustomDarkRed}\texttt{\textbf{#1}}}}

\makeatletter
\newcommand\notsotiny{\@setfontsize\notsotiny\@vipt\@viipt}
\makeatother

\definecolor{LightGray}{gray}{0.978}
  \usepackage{inconsolata}
\usepackage[frozencache,cachedir=.]{minted}
\setminted{%
baselinestretch=.9,%
escapeinside=||,%
fontsize=\scriptsize,%
linenos=off
}
\begin{document}
\newcommand{\eg}{{\sffamily  EG}}

\title{ Streamlining Software Reviews: \\Efficient Predictive Modeling with Minimal Examples}

 \titlerunning{Software Reviews: Efficient Predictive Modeling }        % if too long for running head

\author{Tim Menzies, Andre Lustosa
}


\institute{Tim Menzies, Andre Lustosa\at
              Computer Science, NC State \\ 
              \email{timm@ieee.org, alustos@ncsu.edu}           %  \\
}

\date{Received: date / Accepted: date}

\maketitle

\begin{abstract}
% Various software analytics tasks can be characterized as the 
% struggle to optimize competing goals (e.g. reduce development cost while also reducing the bug count). For such tasks, standard optimization methods have been applied, many of which
% require the evaluation of hundreds to millions of candidate solutions. This is a problem 
% since
% as the evaluation count increases,    the harder it becomes for others to
% understand/ audit/ rebuild/ improve on prior work. 

This paper proposes  a new challenge problem for software analytics.
In the process we shall call ``software review'', a panel of SMEs (subject matter experts) review examples
of software behavior  to recommend how
  to improve that's software's operation.   SME time is usually extremely limited
so, ideally, this panel can complete this optimization task after looking at just a small number of very informative, examples. 

To support this review process, we explore methods that
train a predictive  model to guess if some oracle will like/dislike the next example. 
Such a  predictive model can work with the SMEs to guide them in their exploration of all the examples.
Also, after the panelists leave, that model can be used as an oracle in place of the panel (to handle new examples, while the panelists
are busy, elsewhere).

In 31 case studies (ranging from
from high-level decisions about software processes to low-level decisions about how to
configure video encoding software), we show
that such predictive models  can be built using
as few as 12 to 30 labels.
% Other SE researchers have explored    model building
% using   limited labelling via  unsupervised or semi-supervised learning, or active learning, or zero/few-shot learning.
% But 
To the best of our knowledge, 
this paper's success with only a handful of examples
(and no large language model) is unprecedented.  

% We call this approach LITE since it is  very different to HEAVY methods that (e.g.) need a large language
% model for some generative task
% (such as code translation).
% % As
% % shown in this paper, not all SE tasks are HEAVY. Hence it is
% % prudent to research both LITE and HEAVY methods
% Our hope is that numerous researchers  address the LITE challenge (software review via limited sampling) and discover
% methods that are superior to the one proposed here. Paradoxically, this means
% this paper will be a success if it is quickly superseded.

In accordance with the principles of open science, we offer all our code and data at
\url{https://github.com/timm/ez/tree/Stable-EMSE-paper}
so that others can  repeat/refute/improve these results.

\end{abstract}

%  LAB  would explore (say) 10,000 unlabelled items as follows: \newline
% (a)~use N=4 evalaution to label four   items ; \newline 
% (b)~divide labelled items into ``best'' and ``rest'' using a multi-goal criteria;\newline  (c)~build  a  classifier 
% to find, for unlabelled item, B=like(``best''), R=like(``rest'');\newline 
% (d)~sort, in ascending order, the remaining    9996 items  by -B/R; \newline  
% (e)~label the  top item (so N=N+1), and
% discard the last 20\%   unlabelled items;\newline  (f)~update best and rest with this new item; \newline  (g)~if N$<$25,   loop back  to (b). Else terminate, returning the top item in          ``best''.

\newpage\section{Introduction}\label{intro}

% We explore  models that  predict the effectiveness of an optimizer.  We   show that for many SE domain,
% it takes very little data to train a predictive  model that can guess if an optimizer will like/dislike the next example.  
% Using these
% models, software analytics can stop early (since this predictive model can be
% used for the subsequent reasoning). We call this approach  ``LITE'':
% \begin{quote}
%     {\bf LITE } = Low Input Tuning Enhancement = {\em
%  improve a software system’s behavior
% without having to spend much effort labelling local data. }
% \end{quote}
% %  A similar approach to LITE can be found in    search-based SE~\cite{sarro2023search} where it is assumed that improving
% % system behavior is a multi-objective problem that must  trade-off between competing concerns. But unlike, standard SBSE, 
% LITE 

% An alternative to LITE are HEAVY methods that assume ``the more {\eg}s the better''-- an approach   endorsed by many SE researchers; e.g.~\cite{mcintosh2017fix,rahman2013sample,amasaki2020cross}. 
% Some SE tasks are inherently HEAVY, such as the   creating   a large language model for use on  generative tasks such as code translation~\cite{10.1145/3551349.3559555}. As shown in this paper, not all SE tasks need a HEAVY approach. 

Enabling humans to explore  the important potential behaviors of a software model is an open and important  issue. 
For example,
the more we use AI in SE, the more code will be auto-generated.
The more we auto-generate code, the less time software engineers spend writing and reviewing new code, written by someone or something else (the internals of which they may not understand).
The less we  
understand code, 
the more we will use black-boxes  components,
  where, once a system is assembled, its control settings
are tuned.  
In this scenario, it becomes very  important to reduce
the human effort and CPU effort required for that tuning.

We define ``software review'' as a panel of SMEs (subject matter experts), looking at examples
of   behavior  to recommend how
  to improve   software.   SME time is usually very limited
so, such reviews must  complete after looking at just a small number of very informative examples.  
To support the software review process, we explore methods that
train a predictive  model to guess if some oracle will like/dislike the next example. 
These predictive models   work with   SMEs to guide them as they  explore  the examples.
Afterwards, the models can  handle   new examples, while the panelists
are busy, elsewhere.

This paper   offers an evaluation procedure for software review algorithms.  
Different software review algorithms can be ranked via  {\bf simulation studies} which
(a)~always accept the algorithms next recommendation of ``what to sample''; (b)~then constrains the future search using information taken
from that recommendation.  In such a study, the best software review algorithm is the one that finds the most outstanding examples using
least information.

We present  31 case studies for software review  simulation studies. Also, 
  we   offer a  baseline     review  algorithm called  \verb+lite.py+ that 
addresses the {\em labelling problem}. One way to study some system ``$f$''   
 is to explore examples  \mbox{{\eg}s$\;=(X,Y)$} where $X=f(Y)$.
 Here,   $X$ are the independent variables
and $Y$ are one or more numeric dependent variables (goals). 
{\em Labelling} is the process of computing $Y$ based on    $X$.

As discussed later in this paper, labelling can be slow and/or expensive and/or error prone.
To mitigate this problem, LITE exploits the different costs associated with collecting $X$ and $Y$ values. 
 Specifically, it is often much cheaper to  collect $X$ values     than their associated
$Y$ values. For example,  
while we can quickly count the number of lines of code in a  function,
it is far slower   to design, run, then assess tests that explore that function.
 For other examples where it is much cheaper to find  $X$ than  $Y$, see Table~\ref{slow}.
 
\verb+lite.py+      reasons extensively over the $X$ values
to build models that can make informed guesses about what   $Y$ values are associated with specific $X$ values.
Large data sets, with very few labels,  can then be processed using these models.  
\verb+lite.py+  is less than 25 lines of Python (see Listing~1, in this article). The simplicity of this algorithm is an important feature
since (a)~the simpler  the baseline, the easier it can be   used/changed by newcomers;
(b)~standard TPE implementations are somewhat   complex (e.g. see the 1000s
of lines in   Python's TPE Hyperopt package or the  Watanabe notes~\cite{watanabe2023tree}).

\begin{table}[!t]
{\scriptsize
    \centering
    \begin{tabular}{p{2.2in}|p{2.2in}}
   Find $X$ &  Find associated $Y$ values\\
 \rowcolor{CustomDarkRed}    \textcolor{white}{   It can be very  quick to ...} & \textcolor{white}{ It is a much    slower task  to... }\\ 
   
Mine GitHub to find  all the  distributions of code size, number of dependencies per function, etc.
& Discover (a)how much that software  could be sold on the market or (b)  what is the time required
to build this kind of software\\
% For example, the SZZ algorithm reports what code is associated with a  bug fixing commit. Any number of papers report the complexity of making that determination~\cite{herbold2022problems,ROSA2023111729}.

\rowcolor{gray!20} Count the number of classes in a system. &
Negotiate with an organization permission to fund how much human effort was required to build and maintain that code.\\
       Enumerate many  design options; e.g. 20 binary choices expands to a million options.  &  Check all those options with a   group of human stakeholders.\\
       
 \rowcolor{gray!20}       List the   configuration parameters for some piece of software. & Generate a separate executable for
       each one of those parameter settings, then run those executable through some test suite.\\    
       
      List the   controls of a data miners used in software analytics  (e.g. how many neighbors to use in a k-th nearest neighbor classifier). & Run a grid search looking for 
the best settings for some local data.\\

 \rowcolor{gray!20} Generate test case inputs (e.g.) using some grammar-based fuzzing.& 
Run all all those tests. It can be even slower for a human to check through all those results
looking for   anomalous behavior.\\
 
    \end{tabular}}
    \caption{For $Y=f(X)$, it is often   cheaper to  collect $X$ values     than the 
associated $Y$ values.}
    \label{slow}
\end{table}

% This paper offers new experiments  where  \verb+lite.py+  is applied to the 31 case studies.
%  \verb+lite.py+  significantly out-performs our prior work on SRSAs
% (SWAY~\cite{Chen19} and SNEAK~\cite{lustosa2024learning}). 
% Even for the larger case studies (with 10,000 examples of more), \verb+lite.py+ needs only 30 labels (on average)
% to find results that are statistically indistinguishable from the best known results.

%ozaki2022multiobjective
 
\noindent We  take care to distinguish ``software review'' from other techniques as follows:
\begin{itemize} 
\item
Software review is not software testing. Testing can be an extensive time consuming process to find out as much as possible about all the software. On the other hand, a ``review'' is where SMEs have just a few hours to uncover some significant value-added aspect of a software system.
\item
    Software review is not just few-shot learning  or active learning or the other methods mentioned in our
    literature review.  As discussed in \S\ref{how}, those  methods need large amounts of   background data (to train a large language model) or hundreds to thousands of   labels about the problem-at-and-hand. Software review, on the other hand, assumes access to just a   few dozen labels.
 \item
    Software review is not the same as ``Software 2.0''-- a concept introduced in   Andrej Karpathy's popular medium post\footnote{{\tt https://karpathy.medium.com/software-2-0-a64152b37c35}}.
Software 2.0 a continuous process improvement concept where the core model within a software system
is learned from data. In software 2.0, parameter tuning replaces debugging and  the  training dataset effectively becomes the code base. Our reading of Karpathy's writings is that they assume very large labelled training sets,  processed by deep learning technology, with the goal of automating as much of the process as possible. Here, we seek
algorithms that support  humans ``in the loop'' (by requiring they process very few labels).
\end{itemize}
The rest of this paper is structured as follows:
\begin{itemize}
    \item  In  \S\ref{why}, we
    offer notes on the business case for studying  software review. 
\item 
In   \S\ref{label} we discuss the  labelling problem, and why we must
    not over-label.
 \item In  \S\ref{how}, we discuss   why the problems addressed here  are not solved
    by current research into interactive SBSE, active learning, unsupervised and semi-supervised learning, and zero/few-shot learning.
 \item   \S\ref{details}   describes \verb+lite.py+, a multi-objective TPE~\cite{bergstra11TPE,bergstra2015hyperopt} (tree-structured Parzen estimator\footnote{
Conceptually, TPE explores a tree of candidate estimators. In practice, these algorithms usually do not build a tree but rather allow the sub-tree models to overwrite   parent models.}) that   implements incremental model construction;   i.e. build a model from a handful of labelled {\eg}s (divided into ``best'' and ``rest'')
    then use than model to select what to label next.
\item  \S\ref{experiments} describes  31   SE optimization and configuration case studies. Prior work   needed  thousands to millions of labels for these   studies.
\item \S\ref{results} shows that  \verb+lite.py+'s       significantly improves the default
    behavior seen in those problems, sometimes after  just half a dozen labels.
    To the best of
our knowledge, this is the fewest   labels ever used for these tasks in SE.
\item Finally, in \S\ref{discuss} we discuss threats to validity, future work, and other matters.

  \end{itemize}

\section{The Business Case for Software Review}\label{why}
 % In essence, LITE asks what does it take to change a system 
 % from (a)~its default behaviour  to (b)~some preferred option. This section
 % argues that humans will increasingly need support for this kind of reasoning.

We predict that enhancing our ability to review models will become increasingly important in SE.
In  
``Flaws of policies of requiring human oversight''~\cite{green2022flaws},
Ben Green notes that many recent policies      require humans-in-the-loop to review or   audit   decisions from software models. 
E.g. the  manual of the
(in)famous  COMPAS model notes the algorithm can make mistakes and advises that 
``staff should be encouraged to use their professional judgment and override the computed risk as appropriate''~\cite{northe15}. 

Cognitive theory~\cite{simon1956rational} tells us that such human mistakes
are very common.
  Humans  use heuristic ``cues'' that lead them to the most important parts 
of a model before moving on to their next
task. But when humans review models, they can miss important details. Such cues are essential if humans are to tackle
their busy workloads. That said,  using cues can introduce errors:
   {\em 
   ...people (including experts) are susceptible to ``automation bias'' (involving)  omission errors—failing to take action because the automated system did not provide an alert—and commission error}~\cite{green2022flaws}.
 This means  that   oversight policies   can lead to the reverse of their desired effect  by {\em ``legitimizing the use of   faulty and controversial algorithms without addressing (their fundamental issues''}~\cite{green2022flaws}.

%.  

%Cognitive theory~\cite{simon1956rational} tells us that humans  use heuristic ``cues'' that lets them find    (hopefully)  most important parts of a modelbefore rushing off to their next task.

  Sadly,
there are too many examples of such errors:
\begin{itemize}

\item
Widely-used face recognition software   predicting    gender \& age, has a much
higher error rate for dark-skinned women compared to light-skinned men~\cite{Skin_Bias}.
\item  Amazon's  delivery  software    became   biased against black neighborhoods \cite{Amazon_Bias}. 
\item
  Google Translate  has gender bias. ``She is an engineer, He is a nurse''   translated to    Turkish then
     back to English as ``He is an engineer, She is a nurse'' \cite{Caliskan183}.
  \item The COMPAS model sends   African American defendants to jail   twice as much   as white men~\cite{Machine_Bias}.
  It took years to find and fix this bug
  (and that difference is definitely  a bug since it can be fixed~\cite{fse21} while  maintaining the same levels of recall on actual recidivism).
  \item For  other examples, see~\cite{rudin2019explaining,noble2018algorithms,gebru21}.
\end{itemize}
% 
%That difference is a bug (and we know that is a  bug since it can be fixed-- see PI Menzies' FSE'21 paper~\cite{fse21} that weights training examples to reduce COMPAS's false alarm delta, while  maintaining the same levels of recall on actual recidivism).
% \begin{table}[!b]
% \begin{tabular}{|p{.98\linewidth}|}\hline
% \rowcolor{blue!10}   COMPAS guesses that  black defendants as future criminals   twice as much   as whites~\cite{Machine_Bias}.
% \\
% Widely-used face recognition software   predicting    gender \& age, has a much
% higher error rate for dark-skinned women compared to light-skinned men~\cite{Skin_Bias}.
% \\
% \rowcolor{blue!10} Amazon's   software for same-day delivery    became   biased against black neighborhoods \cite{Amazon_Bias}. 
% \\
%   Google Translate  has gender bias. ``She is an engineer, He is a nurse''   translated to  Turkish then
%   back to  English gives ``He is an engineer, She is a nurse'' \cite{Caliskan183}.\\\hline
% \end{tabular}
% \caption{ Example biases seen in   software decisions. For  other examples, see~\cite{rudin2019explaining,noble2018algorithms,gebru21}.
% }\label{tbl:sigh}
% \end{table}
To forge an effective partnership, humans and artificial intelligence (AI) need to understand each other's strengths and limitations. Software can explore a very large space,
on pre-determined criteria. But humans can offer important and useful insight, but only over
a small number of instances. We are interested in tools like 
\verb+lite.py+
since  they do not make excessive demands of humans reviewing software systems.

\section{  About  Labelling}\label{label}

An alternative to our tools  are   methods that assume ``the more {\eg}s the better''-- an approach   endorsed by many SE researchers; e.g.~\cite{mcintosh2017fix,rahman2013sample,amasaki2020cross}. 
 Such data-heavy methods often require fully supervised learners that have labels  for every example. As discussed in this section,
 ideally we should not need fully labelled data sets.

One complexity with labelling is that it can  be very expensive. 
For example, Tu. et al.~\cite{tu2020better}   studied  714 software projects, including 476K commit files. After an extensive analysis, they  proposed a crowd-sourced cost model for labeling their {\eg}s. Assuming two people checking per commit, that data would need three years of effort to label their {\eg}s.  To place that in context, that effort would consume 40\% of the funds available to a standard three year National Science Foundation grant. To say that another way,
most researchers would be unable to manage that relabelling. 

Another complexity with labelling is that,
  even after spending all that money on labelling,
  the results can be   naive and/or faulty. 
Defect prediction researchers~\cite{catolino2017just, hindle2008large, kamei2012large, kim2008classifying, mockus2000identifying} often  label a commit as "bug-fixing" when the commit text uses words like ``bug, fix, wrong, error, fail, problem, patch''. Vasilescu et al.~\cite{vasilescu2018personnel, vasilescu2015quality} warns that this can be somewhat ad-hoc, particularly if researchers just peek at a few results, then tinker with regular expressions to combine a few keywords.
 Also, when Yu et al.~\cite{9226105} explored labels from prior work exploring technical debt, over 90\% of the "false positives" were incorrectly labelled. Further, there are many reports where data labels error have corrupted the majority of the {\eg}s for security bug labeling~\cite{ 9371393}; for labeling false alarms in static code analysis~\cite{10.1145/3510003.3510214}; or for software code quality~\cite{Shepperd13}.

Since labelling is error-prone, expensive and/or time-consuming, we explore methods that use     as few labels as possible.
One reason
to use minimal labels is that this simplifies the review process described in the last section. Specifically, is is simpler to audit a system when that review means   reflecting over the handful of {\eg}s used used to build the model.

% Other cost models are just as pessimistic:
% \begin{itemize}
%     \item  Valerdi~\cite{valerdi2010heuristics} documentes  the effort associated with getting a panel of experts to agree on the labels of 60 software project effort estimation examples, where each example has 20 attributes. That study required 3*three hour sessions, spread out over one working week. 
%     \item Researchers in the field
%     of repertory grids conduct structured interviews where humans justify attributes and attribute settings for    random subsets of 3 examples, drawn from a larger set. 
%     \begin{itemize}
%     \item
%     Kington~\cite{kington2009defining} reports that it takes humans hour to reflect over  16 examples with 16 attributes.
%     \item
%     Easterby-Smith~\cite{EASTERBYSMITH19803} advises ``keep the grid small. A grid containing ten elements and ten constructs may take two hours to complete. Larger grids may take substantially more time''.
%    \end{itemize}
%    \eind{itemize}
    
For this paper, we say that ``reviewing''  means that some panel of  SMEs (human subject-matter experts) pass
judgement on the observed behavior of a system. We call this ``high-quality'' labelling; i.e. labeling
when it is  required  to carefully justify     labelling decisions (perhaps even to a panel of other experts).  
Such high quality labels can be collected at the rate of   10-20/hour:
\begin{itemize}
\item 
Valerdi~\cite{valerdi2010heuristics} worked with a panel of experts  trying to label  60 software project effort estimation examples, described using   20 attributes. He needed  3*three hour sessions, spread out over a   week. 
\item In iSBSE (interactive search-based software engineering) humans can serve as one of the oracles to guide the search. In their iSBSE research, Takagi et al.~\cite{takagi1998interactive, takagi2000active, takagi2001interactive} argued persuasively  that in order to increase the productivity of SMEs in collecting data for such methods, one should focus on reducing the number of interactions necessary to reach a result. Our SNEAK iSBSE tool~\cite{lustosa2024learning} was designed with  Takagi et al.'s advice in mind. In studies with   human studies and SNEAK, we found we had to  focus on showcasing fewer and smaller interactions at a time. When SNEAK worked with humans, it collected human insight at   the rates suggested above (15-20 labels per hour).
\item
Other research reports similar numbers.   {\em Repertory grids} researchers conduct   interviews where humans justify attributes and attribute settings for    random subsets of 3 examples, drawn from a larger set. 
    Easterby-Smith~\cite{EASTERBYSMITH19803} advises ``keep the (repertory) grid small. A grid containing ten elements and ten constructs may take two hours to complete. Larger grids may take substantially more time''.
Kington~\cite{kington2009defining} agrees, saying  that it takes humans an hour to reflect over  16 examples with 16 attributes using repertory grids.
\end{itemize}
SME expertise is often accessible for just a few hours per month\footnote{Just as an aside, one tactic for getting  more  time  the SMEs
and, hence, more high-quality labels,
is to show   substantive results with the {\eg}s
labelled  so far.  Then,  that expert and the analyst have a business case to take to their manager, saying (e.g.)  ``we are getting some interesting preliminary results. May we get a charge number to bill our time while we work further on this?''. 
But for that to work, some initial study must have obtained interesting results on a small sample
of   {\eg}s.}.
Real-world SMEs  are experts precisely because their expertise is valuable to the organization. 
This means that, by definition,
the SMEs needed for high quality labelling may often   called away to other tasks. Hence we are interested in  scenarios were we are optimizing SE projects using as little as 40 to 80 labels per month: 

\begin{quote} {\bf Task1}:   {\em understand what happens if we can only access 80,40,20 or even as few as 10  labels.}\end{quote}

The results section of this paper will show that
  just labelling (approx) 50 {\eg}s at random can find something that performs better than anything found by a more structured approach. Yet even in those cases, we would not recommend ``try 50 {\eg}s at random'', for the following reason.
  
When it is too slow/expensive to access an labelling oracle (either a human SME or some other source), it is useful to have a model that summarizes what has been learned so far.
 Without such a   model, then if ever we have to optimize new examples, that analysis   will have to pause while we wait for the oracle.   
 But with such a   model, when the oracle is not available, we can still reason about new {\eg}s.
Accordingly:
\begin{quote} {\bf Task2}:   {\em using a few labelled {\eg}s, find a model to
use when it is too slow/expensive to wait for an  oracle.}\end{quote}

Ideally, when an AI models a system, there is space for human insight to guide the reasoning. This is useful for two reasons: (a) humans may have additional background knowledge (unavailable to the algorithm) that can suggest useful solutions;
(b)~humans are more open to using some conclusion if they feel some connection and ownership to
how a conclusion was made. Hence: 

\begin{quote} {\bf Task3}:   {\em when exploring a a few   {\eg}s, allow some method
for humans to (optionally) have some control over   the algorithm.}\end{quote}

% XXX task3 explanation

% ### task4 accept input from human

\section{Related Work}\label{how} 

Much    research has explored    model building
using a few labelled {\eg}s.  But as we shall see in this section,
(with some caveats) those method are rarely applied on as little as 80,40,20 or 10 labels.

\subsection{Zero/Few-Shot Learning, Active/Semi-supervised Learning, and Others}\label{others}

There are two kinds of algorithms that make decisions about software without   new labels from the test domain.
 \underline{\bf  Zero-shot learners} use  the  background knowledge of a large language model to make decisions without needing new labels~\cite{alhoshan2022zero}.
 Zero-shot learners  works in domains where there exists an
  appropriate large language model (which is {\em not}
  the situation for our case studies).
On the other hand, \underline{\bf unsupervised learners} use a domain heuristic to classify examples by 
 peeking at the independent variables:
 E.g. Nam et al. successfully predict for defective modules by looking for classes that are unusually large~\cite{nam2015clami}. Such domain heuristics are not known for our case studies.

 Another technique for labelling less is to recognize  some system condition
 under which an {\eg}
can be unequivocally labelled as ``fail''; e.g. (a)~if a test generates a core dump or (b)~some metamorphic predicate~\cite{chen2018metamorphic} argues that, say, ``small changes to inputs should not cause large changes to 
output''.
 The value of this approach is that such a domain-general labelling
oracle  might be relevant to other applications.
But on the flip side, the more general the test oracle, the less it connects
to the specific goals of the local users.
 For example, in the case studies shown below, we test for user-supplied 
 local goals such as ``if we implement this requirement next, then some other team will not stand idle waiting for some other function we were meant to implement''.
 
 Another class of algorithms uses less labels by, say,    labelling just a few {\eg}s,  then propagating those values to other near-by {\eg}s.
Such  \underline{\bf semi-supervised} learners have successfully reasoned over 10,000s   records, after labelling just 1 to 2.5\% percent of the {\eg}s~\cite{10109333,majumder2024less}.
While a useful approach, in the studies we have seen~\cite{10109333,majumder2024less},``1 to 2.5\% 
of the data'' still means  
100s to 1000s  of labels. 

Human-in-the-loop   \underline{\bf  active learners}  
minimize human labelling effort by updating their models
each time a human offers an opinion on an {\eg}.
Even with these tools, the labelling effort can be excessive.
Yu et al.'s active learners need 20\% of the data
labelled, which for (e.g.) 28,750 files in Firefox means labelling 5,750 {\eg}s~\cite{yu2019fast2}. Similar values of 10 to 20\% labelling
have been reported by other active learning researchers in SE:
see   Wu et.al~\cite{WU2021106530}.

Researchers in  \underline{\bf  interactive-search-based SE} explore human-in-the-loop optimization where humans serve as the oracle that judges  if one {\eg} is better than another.
When humans are asked too many questions,
they start making mistakes due to  {\em cognitive fatigue}~\cite{shackelford2007implementation}.
One solution  is to learn a {\em surrogate model} from a small sample of {\eg}s, then querying the surrogate for labels~\cite{araujo2017architecture}.  But even with technologies like surrogates, our reading of the iSBSE
research~\cite{lin2016interactive,amal2014use,araujo2017architecture} is that community still assumes  hundreds of examples can be labelled.

In the SE literature, one area close to this paper
is
\underline{\bf few-shot learning}.
This is a technique that uses 
a few dozen  labeled {\eg}s (sometimes as few as ten)   
to convert a general large  language models into some specific tool; e.g. 
 parsing   test  case output~\cite{le2023log} or translating functions into English~\cite{10.1145/3551349.3559555}. But
 like zero-shot learning, few-shot learning
 only works where, in the domain being explored, there exists
 a functioning large-language model. For the case studies presented below, we are unaware of such models.

\subsection{Prior Work with SWAY and SNEAK}\label{sway}
This paper evolved our prior work on  SWAY~\cite{Chen19} and SNEAK~\cite{lustosa2024learning} algorithms, which recursively divided the data based on the $X$ attributes. 
Some assessment method was applied to prune the 
worst half. 
 In this paper, we assess using  the
{\em distance to heaven measure} assessment measure (hereafter {\em d2h}) of Figure~\ref{d2h}.
SWAY   then recursively applies itself on the surviving half.  In this way, SWAY and SNEAK would try to explore
$N$ examples using $O(\log_2(N))$ labels.

One drawback with SWAY is that it is    a ``greedy search`` that pruned half the data
immediately after making any split.   This approach is prone to early, possibly suboptimal decisions because it might eliminate promising examples without sufficient consideration. SNEAK aimed to refine SWAY's greedy strategy by initially bi-clustering all the {\eg}s   without any elimination. After building the entire tree, SNEAK reviewed all internal nodes to select a ``large'' and ``most informative'' node that was also ``easiest to ask'' where:\begin{itemize}\item ``Large'' indicates preference for nodes with the most examples.\item ``Most informative'' refers to nodes whose sub-trees significantly lower the entropy of the independent $X$ attributes.\item Nodes that are ``easiest to ask'' about have minimal differences, allowing for a concise presentation of these differences.\end{itemize}
SNEAK divided the data at this node; the less valuable half was eliminated according to the criteria in Equation~\ref{d2h}; and the procedure was replicated with the remaining data.
To say all this another way:\begin{itemize}\item SWAY's aggressive  strategy   pruned the data at the earliest opportunity;\item   SNEAK postponed pruning till it can considered  more  possibilities.\end{itemize}Initial tests with SNEAK, based on three datasets, showed encouraging results~\cite{lustosa2024learning}. Nevertheless, after evaluating the 31 datasets discussed in this report, we must retract our earlier endorsement of SNEAK as SWAY outperforms it. Moreover, \verb+lite.py+'s predictive modeling approach, which utilizes a non-greedy search that incorporates extensive global reflection, decisively outperforms both SWAY and SNEAK.

 For information on SNEAK (which is being discontinued due to subpar performance), consult ~\cite{lustosa2024learning}.
 For more information on SWAY and \verb+lite.py+, see below.

% \subsection{DODGE}

% The  
%    DODGE hyperparameter optimizer~\cite{agrawal2019dodge}. 
%  DODGE was a tabu search that encouraged an optimizer to avoid new {\eg}s   that generated solutions that were too close
% to old solutions.   
% DODGE   only needed 30 {\eg}s
% to configure learners for  for a range of SE tasks. 

% To our shame,  DODGE was never compared to  a purely random selection of 30 {\eg}s until several years after its publication. While we cannot not defend that oversight, we note that ``compare with random'' does not appear in SE's lists of ``best practices'' for this kind of research:
% \begin{itemize}
% \item
% Such comparisons are not mentioned in {\S}9 of the    Arcuri\& Briand text~\cite{10.1145/1985793.1985795}.
% \item
% In the Empirical SE standards docs at 
% \url{https://github.com/acmsigsoft/EmpiricalStandards}, footnote 7 of 
% docs/standards/OptimizationStudies.md suggests that such random comparisons are depreciated.
% \end{itemize}
% In any case, 
% DODGE is not reported in this since the ``compare with 30 random condigurations'' results reported below typically performed as well as DODGE.  

\section{    Algorithmic Details} \label{details}
This section offers details on the alaogithms explored in our experiments.

\subsection{Random(N)}
In his textbook {\em Empirical Methods for Artificial Intelligence}, Paul Cohen~\cite{Cohen:1995} advises that seemingly
sophisticated algorithms should be compared against some simplest alternative-- if for no other
reason that this offers baseline performance values.  Hence, the experiments of this paper will
compare our proposed algorithms against a  simple random choice approach called random(N).

Random(N) runs as follows.
\begin{itemize} 
\item Select   $N$ examples at random,
\item Label   all $N$ items,  sort them by some criteria. We will use the {\em distance to heaven} measure of Table~\ref{d2h}.
\item From that sort, return the top-most example.
\end{itemize}

\begin{table}[!t]
  {\small   \begin{tabular}{|p{.965\linewidth}|}\hline
    After
labelling, all our {\eg}s have labels for each  goal.
Further, each goal
has a range \mbox{ {\em lo .. hi}} so goal values can be normalized to   0..1. Each goal 
has a  {\em heaven}  of  0,1 if 
 that goal is to be  minimize or maximized  (respectively). We say an example has a distance to heaven ({\em d2h}) equal to the Euclidean distance of the goal values to a mythical best value. For example, if we want to minimize
\textit{ bugs} and \textit{features}, and these have the  range 0 to 100, then for an example with 30 \textit{bugs} and 80 \textit{features}: 
\begin{equation}\label{d2h}
\begin{array}{rl}
\mathit{d2h}((X,Y)) & = \mathit{dist}(\mathit{heaven}, \mathit{norm}(Y))\\ 
& =  \frac{\sqrt{ (0-.3)^2 +  (1-.8)^2}}{\sqrt{2}}\\
& = 0.26
\end{array}
\end{equation}
We divide the distance by the $\sqrt{|\mathit{goals}|}$, to ensure    \textit{d2h} has a range zero to one. 
Note that {\em smaller} values of \textit{d2h} are \textit{better} and so SWAY and SNEAK and \verb+lite,py+ try to find subsets of the examples with very small distances to heaven.\\\hline
\end{tabular}}
\caption{Distance to heaven.}\label{d2h}
\end{table}

Random(N) is very simple to code and has some   reliable mathematical properties. 
 Hamlet~\cite{hamlet1987probable} argues that after $n$ random trials, events occurring at probability $p$ can be observed at confidence $C=1-(1-p)^n$. Solving this equation for $n$,   we can derive the following relationship 
between $n,C,p$:
\begin{equation}\label{hamlet}
n(C,p) =   \log(1-C)/\log(1- p)  
\end{equation}
Jacob Cohen~\cite{cohen2013statistical}    argued
that two events are statistically indistinguishable  if they differ by less than .35 of the standard deviation $\sigma$.
For solutions whose independent variables come from   a  normal population running from $-3\sigma \le x \le 3\sigma$,
this means there are 
  $(6\sigma)/(.35\sigma)\approx17$ equivalence classes of distinguishable solutions; i.e. a defensible  estimate for $p$
   is $p=1/17$. Hence, to be   $C=95\%$ confident of  finding  examples at probability $p=1/17$ that are statistically indistinguishable from the best solution, we need
\[\log(1-.95)/\log(1-1/17) \approx 49\;\mathit{labels}\]
As seen below, this estimate is remarkably close to what is observed in practice.
But
while a useful baseline,
this paper will not endorse random(N) for software review:
\begin{itemize}
    \item
 
Random(N) does not know how to use SME knowledge which, as   argued in \S\ref{label},
is a desirable property for software review.
On the other hand, the \verb+lite.py+ method described below could check and change its conclusions with a human at each step in its sampling loop.
\item Random(N) does not produce a model-- which means if we ran it twice,  a week apart,
    then    random(N) would need just as many samples in the  second session as in the first. On the other hand, in the second session, the  \verb+lite.py+ method described in the next section could reuse
    (without any further labelling) the model acquired from the first session.   
\end{itemize}

\subsection{{\tt lite.py}}

\verb+lite.py+ builds a model for the ``best'' and ``rest'' {\eg}s labelled so far. It then (a)~guesses
what unlabelled example might be most interesting, (b)~labels it,  (c)~rebuilds the ``best,rest'' model;
then (d)~loops back to guess again.  To define ``intesting'', we use the
 {\em distance
to heaven calculation} of Table~\ref{d2h}.

To illustrate how \verb+lite.py+ works, imagine we have 10,000 cars to label,  some of which have the   $N=9$ labels of Table~\ref{36}. 
For this example, we will assume that:
\begin{itemize}
    \item 
We prefer lighter cars since they use less raw materials and are 
  cheaper to build. Hence, we want to minimize car weight (see the ``-'' sign after ``Lbs'' on row one of  Table~\ref{36}).
  \item The other goals are to  maximize car acceleration and miles per gallon (see the ``+'' signs on those goals in  row one of Table~\ref{36}).  
  \end{itemize}
  After normalizing the goals to the range 0..1 
  (using their minimum and maximum values)
  `heaven'' is the point 
    (0,1,1) for (``Lbs,Acc,Mpg'') respectively.
  With knowledge of  heaven , the   distance to heaven ({\em d2h}) calculator
of Equation~\ref{d2h}, can sort these {\eg}s. These can then be divided  into (say) 
$\sqrt{N}=3$ ``best'' examples and the remaining ``rest''.

Using these divisions, \verb+lite.py+ builds a two-class Bayes classifier. This predictive model   computes
the likelihood $b,r$ that  some unlabelled {\eg} belongs to class ``best'' or ``rest''. 
Using $b,r$, there are several policies for  guessing what  {\eg} should be labelled next. Yu et al.~\cite{8883076} says that the most informative {\eg} is the one
that most challenges existing ideas.   Under this assumption, the next best example
is that one that is most controversial i.e. $b$ and $r$ are both large and very similar:  \begin{equation}\label{uncertain}\mathit{uncertain}=(b+r)/abs(b-r)\end{equation}  
An opposite approach, used with   success by Menzies et al. in  
SE applications~\cite{Me07,menzies10dp}, is to seek unlabelled example that most confirm   current beliefs.
This would be the example that is mostly likely to be ``best'' and least likely to be ``rest''; i.e.
\begin{equation}\label{certain}\mathit{certain}= b / r\end{equation}

\begin{table}[!t]

\begin{center}
{\scriptsize \begin{tabular}{|rrrr|rrr|r}
\multicolumn{4}{c|}{ X  }&\multicolumn{3}{c}{ Y labels }\\
\multicolumn{1}{c}{Clndrs} & Volume &   Model & origin & Lbs- & Acc+ & \multicolumn{1}{c}{Mpg+} &  \\\cline{1-7}
 4& 97&   82& 2& 2130& 24.6& 40&\\
4& 96&   72& 2& 2189& 18& 30& 3 ``best''\\
 4& 140&  74& 1& 2542& 17& 30&\\\hline
 4& 119&   78& 3& 2300& 14.7& 30&\\
 8& 260&   79& 1& 3420& 22.2& 20&\\
 4& 134&   78& 3& 2515& 14.8& 20& 6 ``rest''\\
6& 231&  78& 1& 3380& 15.8& 20&\\
8& 302&   77& 1& 4295& 14.9& 20&\\
8& 351&  71& 1& 4154& 13.5& 10\\\cline{1-7}
\multicolumn{4}{c|}{independents  }&\multicolumn{3}{c}{   dependents   (a.k.a. goals)}\\
\end{tabular}}
\end{center}
\caption{$N=9$  {\eg}s, sorted by Equation~\ref{d2h}. Divided into   $\sqrt{N}=3$ ``best'' and 6 ``rest''.   ``Best'' cars (at top of table)    weigh less and   have most
acceleration and miles per hour. }\label{36}
\end{table}

\begin{listing} 
\begin{minted}{python}
def label(row): ... # esnure dependent vars are computed from independent vars; return the row 

class the # stores the settings for this system
    start = 4   # intiially label 4 examples  |\six|
    halt = 16   # label up to 16 more examples  |\six|
    best = 0.5  # after sorting labelled items, sepearte the top n**'best'   
    upper= 0.8  # after sorting unlabelled items, keep top (say) 80%
    m,k  = 2,1  # naive bayes low frequency management   |\twelve|
    # SWAY settings (used in Listing |\ref{listing:sway}|)
    Far=0.95    # when look for distant EGs, avoid outliers. stop at 95%
    Half=256    # when look for distant EGs, search 256 EGs, at random
    Stop=0.5    # SWAY terminates on leaves of size n**Stop
    Dive=50     # SWAY2' first round terminates on leaves of size 50
    deeper=4    # SWAY2's second round terminates on leaves of size 4
  
class COL:   # summarizes one columns of the examples
    column = 0                # position of this this numeric column
    n      = 0                # number of items added to COL
    def add(self,  x):   ...  # increment 'n', add x to this distribution 
    def dist(self, x,y): ...  # use the Aha equation from |\S\ref{sway}| to compute distance. |\three|
    def like(self, x):   ...  # prob that `x' is in this distribution
  
class NUM(COL): # summarizes numeric columns |\one|
    mu = 0
    sd = 0
    heaven = 0 or 1           # '0' if minimizing, '1' if maximizing
    def like(self, x: float): # assume a gaussian; to avoid division errors, uses |$\epsilon=10^{-64}$| |\four|
        return |$\frac{1}{(\sigma+\epsilon)\sqrt{2\pi}} \exp\left(-(x-\mu)^2/(2\sigma^2 +\epsilon)\right)$| 

class SYM(COL): # summarizes non-numeric columns   |\two|
    has = dict[str,any]
    def like(self,x,m:int,prior:float): return (self.has.get(x, 0) + m*prior) / (self.n+m)|\four| |\twelve|
  
class DATA: # stores the rows and the column summaries 
    rows = list[list]       # list of examples
    x    = list[NUM | SYM]  # columns for independent attributes
    y    = list[NUM]        # columns for depedent goal attributes
    def clone(self, rows=[], order:bool) # make new DATA,same headers, new rows (maybe sorted) 
    def d2h(self, row)                   # return the distance of row to the heaven point.
    def sort(self,rows=[]):  self.rows = sorted(rows, self.d2h)  
    def like(data:DATA, row:list, nall:init, nh:int): # log likelihood of row is in data 
        prior = (len(data.rows) + the.k) / (nall + the.k*nh) |\twelve|
        tmp = [col.like(row[col.colulmn], the.m, prior) for col in data.x]  |\twelve|
        return sum(math.log(x) for x in tmp + [prior])

def predictiveModeling(data: DATA, 
                       guess=lambda B,R: B-R): |\eleven| # really B/R since applied to log likelihoods
    def like(row,data2): # Return log likelihood of row belonging to data2
      return data2.like(row, len(data2.rows), 2) |\five|
    def acquire(best, rest, rows): # Sort rows best to rest, keep the first (say) 80%"
      chop = int(len(rows) * the.upper)
      return sorted(rows, key=lambda r: -guess(like(r,best),like(r,rest)))[:chop] |\nine|
      
    random.shuffle(data.rows)
    done, todo = [label(row) for row in data.rows[:the.start]], data.rows[the.start:] |\seven|
    data1 = data.clone(done, order=True) |\thirteen|
    for i in range(the.halt):
      if len(todo) < 3: break # todo is too small
      n = int(len(done)**the.best + .5)
      top,*todo = acquire(data.clone(data1.rows[:n]), data.clone(data1.rows[n:]), todo) |\eight|
      done.append(label(top)) |\ten|
      data1 = data.clone(done, order=True) |\ten| 
    return data1.rows[0],len(data1.rows)
\end{minted}
\caption{\RED{predictiveModeling}  guesses if   unlabelled {\eg}s   are   ``best'' or ``rest''.}
\label{listing:1}
\end{listing}

The experiments of this paper will try both approaches. Interestingly, while the  literature usually recommends some form of uncertainty sampling (e.g.~\cite{muller22}), we did not find that it worked best.

Listing~\ref{listing:1} shows  \verb+lite.py+. The upper half of this code is a generic data model
that loads in data from (say) a csv file, then summarizes that code in some column headers. The last
22 lines show the inference that is specific to \verb+lite,py+.
Also of note is   
the \RED{label} function (on line one)  that look at the independent variables, and compute the dependent variables.
If the {\eg} already has labels, then \RED{label} just returns the labelled {\eg}, unchanged.

All the {\eg}s are stored in the {rows}
attribute of  a
 DATA object. The columns of those rows are summarized in   COLumn objects which specialize
 to NUMs {\one} and SYMs {\two} (for numeric and symbolic columns). The \RED{dist} method of COLumns can calculate the distance between
 two values using the  Aha equation {\three}
 from  \S\ref{sway} (which is   used later when we discuss SWAY). Also, the \RED{like} method knows how
 to return  the likelihood that a number belongs to a distribution {\four}. These methods are called 
 when DATA objects need to compute $b,r$ for an unlabelled row {\five}.

In the \RED{predictiveModeling} function of Listing~\ref{listing:1}, {\em done} and {\em todo} hold the labelled and unlabelled rows.
The creation and processing of these variables are 
controlled by the parameters {\em the.start} and {\em the.halt} (defined at the top {\six} of 
Listing~\ref{listing:1}).
Initially, we label just {\em the.start=4}  rows {\seven}.  We then loop {\em the.halt=16}
times where the   $n$ currently labelled
{\eg}s from {\em done} are divided into $\sqrt{n}$ {\em best} and {\em rest} {\eight}.
The remaining unlabelled {\em todo} items are then sorted by (e.g.) $b/r$, after which we (a)~forget
the worst ones {\nine}; (b)~extract the best one for labelling; (c) add it to {\em done}, which is  resorted  {\ten};  and then the process loops.

 After collecting a fixed number  of labels,  \RED{predictiveModeling} terminates 
and returns (a)~the top-most sorted row and (b)~the number of labels (equal to the size of
the {\em done} list). As per {\bf Task2} (from \S\ref{label}), the model stored in {\em data1}
is also available for caching and reuse if ever new {\eg} come along.

Some other points are worth noting:
\begin{enumerate}
\item Recall that 
Equations~\ref{uncertain} and ~\ref{certain} were different policies to decide what {\eg}s to label next. We pass these policies 
as parameters to \RED{predictiveModeling} so that the same code base can implement both policies {\eleven}
\item  $m,k$  are small offsets
to handle low attribute counts or low class frequencies {\twelve}.  
 Webb et al.~\cite{yang2002comparative} recommend these since low frequencies can confuse Bayes classifiers. 
\item  \RED{predictiveModeling} does not confuse its train and test data.  
The {\em data} variable (passed in as the its first argument) is only used to supply a list of unlabelled rows, plus some meta-information on
the column headers. This meta knowledge is needed so the function can  \RED{clone} a new DATA instances that know how many columns to hold, and which are numeric ir goals or otherwise. The first time this function uses specific values for {\em rows} is when we create {\em data1} from just the first
{\em the.init} rows {\thirteen}.
\item  Every time this code sees a new {\eg},
the entire Bayes model is torn down and rebuilt. This is not slow since (a)~updating a Bayes classifier is very fast (only requires updating some frequency counts) and (b)~these models  are built (and re-built) using very few  {\eg}s.
\end{enumerate}

\renewcommand*\circled[1]{\tikz[baseline=(char.base)]{%
            \node[minimum width=1pt, shape=circle,fill=alizarin,inner sep=1pt] (char) {{\footnotesize \textcolor{white}{#1}}};}}

% \newcommand*\circled[1]{\tikz[baseline=(char.base)]{
%             \node[minimum width=1pt, shape=circle,fill=alizarin,inner %sep=1pt] (char) {{\footnotesize \textcolor{white}{#1}}};}}

% def sway2(data:DATA): # SWAY twice. Second time, use rows found the first time
%     random.shuffle(data.rows);
%     best1, rest1, labels1, __   = sway(data.rows, the.Dive)
%     best2, rest2, labels2, last = sway(best1, the.deeper)
%     return  best2, rest1+rest2, labels1 + labels2, last
            
\subsection{{\tt SWAY}}\label{swaycode}

SWAY, shown in Listing~\ref{listing:sway}, uses the DATA, COL, NUM, SYM objects from  
Listing~\ref{listing:1}.
  SWAY  
   implements optimization as a binary chop through a space of {\eg}s
   clustered on the independent attributes.
   SWAY was inspired by   algorithms that clustered data using  PCA (principle components analysis~\cite{pearson1901principal}).
The advantage of PCA is that some data sets are better described by synthesized attributes that combine the influences of many raw attributes
into some new dimension.
But standard PCA can only handle numeric data so SWAY combined the Aha et al.~\cite{aha91} distance metric\footnote{ Aha et al.~\cite{aha91}  suggest that points $(X_1,Y_1), (X_2,Y_2)$ are
separated by $\mathit{dist}(X_1,X_2)= \left(\sum_i \Delta(X_1^{'},X_2^{'})^2\right)^{0.5}$. For discrete values, 
\mbox{$\Delta(a,b)=0\;\mathit{if}\;a==b\;\mathit{else}
\;1$}.
For numeric values,
$\Delta(a,b) = 
\mathit{abs}(\hat{a} - \hat{b})$ where $\hat{a}$ is $a$   normalized to the range 0..1 for min..max.
For missing values, we use values for 
$a,b$ that maximizes the computed distance.
}
with the FASTMAP heuristic~\cite{faloutsos1995fastmap} to recursively bi-cluster {\eg}s using the independent attributes. Formally, SWAY belongs to the family of Nystr\"om algorithms that approximate PCA, by various methods~\cite{platt2005fastmap}.

\begin{listing}[!t]
\begin{minted}{python}
#class DATA continued
  def near(self, row1, rows=None): # Return rows, sorted by dist to row1
    return sorted(rows or self.rows, key=lambda row2: self.dist(row1,row2))
    
  def faraway(self, rows, sortp=False, last=None): # Get 2 distant items, maybe reusing last 
    n     = int(len(rows) * the.Far) |\four|
    left  = last or self.near(random.choice(rows),rows)[n] |\two| |\nine|
    right = self.near(left,rows)[n] |\three|
    if sortp and self.d2h(label(right)) < self.d2h(label(left)): left,right = right,left
    return left, right

  def half(self, rows, sortp=False, last=None): # divide data by distance to 2 distant egs 
    def dist(r1,r2): return self.dist(r1, r2)
    def proj(row)  : return (dist(row,left)**2 + C**2 - dist(row,right)**2)/(2*C) |\seven|
    left,right = self.faraway(random.choices(rows,  |\one|
                              k = min(the.Half, len(rows))),  |\five|
                             sortp=sortp, last=last)
    lefts,rights,C = [],[], dist(left,right)
    for n,row in enumerate(sorted(rows, key=proj)):
      (lefts if n < len(rows)/2 else rights).append(row) |\eight|
    return lefts, rights, left

def sway(data:DATA, rows=None,stop=None,rest=None,labels=1, last=None): # Recurse to best half
    rows = rows or data.rows
    stop = stop or 2*len(rows)**the.Stop
    rest = rest or []
    if len(rows) > stop: |\ten|
      lefts,rights,left  = data.half(rows, True, last)
      return sway(lefts, stop, rest+rights, labels+1, left)
    else:
      return rows, rest, labels, last
\end{minted}
\caption{\RED{sway}'s  recursive bi-clustering.}
\label{listing:sway}
\end{listing}
 At each level of the recursion, SWAY uses  the   Aha distance measure
 from \S\ref{sway} to find two faraway examples {\em left, right} {\one}. 
 These  are found in linear time using the FASTMAP heuristic~\cite{faloutsos1995fastmap}:
 \begin{itemize}
 \item  {\em left} is found by looking for something ``distant'' to any random {eg}; {\two}
 \item  {\em right} is found by looking for something ``distant'' to {\em left}; {\three}
 \item  In order to avoid extreme outliers, ``distant''  is defined to be  {\em the.Far=0.95} times max distance ; {\four}
 \item To speed   the code, we   seek ``distant'' {\eg}s in just    {\em the.Half=256} points;  {\five}
 \item We sort {\em left,right} via distance to heaven (so {\em left} is always best). {\six}
     This is the only point where SWAY looks at {\eg} labels. That is, SWAY never needs to  label more than two {\eg}s per
     level.  
 \item Using the cosine rule\footnote{If $X$ has distance $a,b$ to points $A,B$, then $X$ projects to a line $\overline{AB}$ of length $c$ 
 at a point $x=(a^2+c^2-b^2)/(2c)$.}     {\eg}s are projected onto a line from {\em left} to {\em right}; {\seven}
 \item {\eg}s are then divided at the median point between {\em left} and {\em right}. {\eight}
 \end{itemize}
  SWAY then recurses on the examples nearest the ``best'' of {\em left} and {\em right}.
 On recursion, when looking for two more distant points, SWAY 
 reuses (without relabelling) the  last ``left''   {\eg} from the  parent. This means that SWAY needs two evaluations for the root split (since, there,  there is no parent cluster), then one new labelling for each level of recursion {\nine}.

 SWAY terminates
when the leaf classes contain less than some user-specified point; e.g. $n^{the.Min=0.5}$ {\ten}. 
On termination, it returns the last ever ``best'' example found by the algorithm.
For clustering $N=10,000$ examples into bins of size $\sqrt{N}=100$, SWAY requires nine labels.

\section{Experiments}\label{experiments}
The rest of this paper uses simulations studies to compare and rank algorithms for supporting
software review.

These experiments will explore   several research questions:
 \begin{itemize}
 \item
{\bf RQ1:} What are the least number of labels we should use?
\item 
{\bf RQ2:} What are the most number of labels we should use?
\item
{\bf RQ3}: What are the merits of purely random labelling?
\item
{\bf RQ4:} For non-random labelling, what policy should be used? 
\end{itemize}

\subsection{Methods}
 
\subsubsection{Algorithms}

  This paper will explore the SWAY, SNEAK, \verb+lite.py+ and random(N) algorithms defined above.
  We explore  SNEAK and SWAY since we need to check if the proposed new method
  \verb+lite.py+ is any better than our prior work.   We explore random(N)
  since Cohen~\cite{Cohen:1995} advises
  that
  seemingly sophisticated algorithms should be compared against
some simpler alternative.

   We do not compare
  against few-shot learning since  we are unaware of LLMs relevant to to the varied nature of our case studies. We do not compare against the active learning and semi-supervised learning algorithms of since, as discussed in \S\ref{others}, these
  algorithms routinely require 100s of labels (or more).

\subsubsection{Data}

Our simulation studies are conducted on the case study data of Table~\ref{data}. Other optimization studies rely on  models to generate data, incrementally, using what ever has been learned up to this moment within the optimization.  
Such models can be turned into our kind of data input by running a very large set of random model inputs
(which would be the $X$ values) then caching
model output (which would be the $Y$ values).
Previously, Chen et al. have shown that such a
pre-compute and cache strategy can generate optimizations competitive to incremental model-based generation~\cite{chen16}.

Our data comes from prior work that 
explored optimization/configuration problems in SE~\cite{nair18tse,Nair2016,chen18,lustosa2024learning,Me07,menzies2009avoid,green2009understanding,port2008using}. All the items in   Table~\ref{data}.
have the    property that some prior work has analysed that data using algorithms that required thousands to millions
of labels. For example, when we first analyzed the flight, ground, osp, osp2 data sets (listed towards the bottom 
of Table~\ref{data}), we used a simulated annealing algorithm that   required 100,000 labels per  data set~\cite{Me07}.
 
 \begin{table}[!t]
 \begin{adjustbox}{width = \textwidth, center}
{\scriptsize\begin{tabular}{p{3cm}p{1.7cm}rc|cr|rr}
\multicolumn{4}{c|}{~}&\multicolumn{2}{c|}{X}&\multicolumn{2}{c}{Y}\\ 
\rowcolor{CustomDarkRed}\textcolor{white}{group}&\textcolor{white}{name}&\textcolor{white}{rows}&\textcolor{white}{cols}&\textcolor{white}{NUMs}&\textcolor{white}{SYMs}&\textcolor{white}{minimize}&\textcolor{white}{maximize}\\
Config of misc cloud&SS-A & 1343     &  5 & 3 & 0 & 1 & 1    \\
utils on Apache Storm     &SS-B & 206 &      5 & 3 & 0 & 2 & 0    \\
&SS-C & 1512 &       5 & 3 & 0 & 1 & 1    \\
 &SS-D & 196 &      5 & 3 & 0 & 1 & 1    \\
&SS-E & 756 &      5 & 3 & 0 & 1 & 1    \\
    &SS-F & 196 &       5 & 3 & 0 & 1 & 1    \\
&SS-G & 196 &      5 & 3 & 0 & 1 & 1    \\
&SS-H & 259 &     6 & 4 & 0 & 2 & 0    \\
&SS-I & 1080 &      7 & 5 & 0 & 1 & 1    \\
&SS-J & 3840 &      8 & 6 & 0 & 1 & 1    \\
&SS-K & 2880 &      8 & 6 & 0 & 1 & 1    \\ 
\rowcolor{gray!20} Compiler configuration&SS-L & 1023 &       13 & 11 & 0 & 2 & 0   \\ 
% &SS-M & 239360 & se & auto & note & 15 & 13 & 1 & 1 & 0    \\
% &SS-N & 53662 & se & auto & note & 19 & 17 & 0 & 2 & 0    \\
% &SS-O & 65424 & se & auto & note & 61 & 59 & 0 & 2 & 0   \\
 
  Predict GitHub health  &   closedPR-0 & 10000  &  8 & 4 & 1 & 1 & 2   \\
   indicators in 12 mths&  closedPR-1 & 10000 &    8 & 4 & 1 & 1 & 2   \\ 
     &   closedIssues-0 & 10000  &  8 & 4 & 1 & 1 & 2   \\
  &  closedIssues-1 & 10000 &    8 & 4 & 1 & 1 & 2   \\ 
     &   commits-0 & 10000  &  8 & 4 & 1 & 1 & 2   \\
   &  commits-1 & 10000 &    8 & 4 & 1 & 1 & 2   \\

\rowcolor{gray!20} Waterfall software                        &china & 499 &     19 & 16 & 0 & 1 & 0   \\
\rowcolor{gray!20} predictors of (e.g.)                           &coc1000 & 1000 &     25 & 0 & 17 & 4 & 1   \\
\rowcolor{gray!20} effort, defects, risks          &nasa93dem & 93 &      29 & 1 & 22 & 2 & 1   \\
 \rowcolor{gray!20}                       &flight & 10000 &     31 & 1 & 22 & 4 & 0  \\
 \rowcolor{gray!20}                       &ground & 10000 &     31 & 1 & 22 & 4 & 0    \\
  \rowcolor{gray!20}                        &osp & 10000 &      31 & 1 & 22 & 4 & 0    \\
  \rowcolor{gray!20}                         &osp2 & 10000 &      31 & 1 & 22 & 4 & 0    \\ 
 Agile development &pom3a & 500 &       12 & 9 & 0 & 3 & 0   \\
  predictors of (e.g.) &pom3b & 500 &      12 & 9 & 0 & 3 & 0   \\
  completion  rate    &pom3c & 500 &      12 & 9 & 0 & 3 & 0    \\
  idle rates,dev cost                &pom3d & 500 &      12 & 9 & 0 & 3 & 0   \\
\rowcolor{gray!20} Car   design&auto93 & 398       &   8 & 3 & 1 & 1 & 2    \\
\rowcolor{gray!20} WineQuality& wine & 1599     &  12 & 10 & 0 & 1 & 1    
\end{tabular}}
\end{adjustbox}
\caption{Thirty one multi-objective case studies (mostly from SE).}\label{data}
\end{table}

All the data sets of Table~\ref{data} have the   structure of 
Table~\ref{36}; i.e. they are csv files whose first row names each columns.
Names stating in upper case denote numeric columns and all other columns are symbolic.
Names ending in ``+'' or ``-'' denote the $Y$  goals to be minimized or maximized. 
For testing purposes, our data has all the
  $X,Y$ values filled in.
  The first time a row's  $Y$ value is accessed, the \RED{label} function of
  Listing~1 is triggered to increment a ``number of labellings'' counter.

The shading of Table~\ref{data} divides that material into several groups. The groups
names ``SS-*'' come from the software configuration literature~\cite{Nair2016}. This data was running
software configured in different ways (selected at random), then collecting
various performance statistics (runtimes, query times, energy usage, etc). 

The next group in Table~\ref{data}  comes from the hyperparameter optimization literature~\cite{lustosa2024learning}.
The ``*-0, *-1'' data sets show results where random forest regression algorithms were configured to predict for 
number of (a)~commits or (b)~closed issues or (c)~close pull requests in 12 months time in open source projects
housed at GitHub. The $Y$ values of these data sets show the results of the predictions after certain hyperparameters were applied
to the random forests (which, in turn, were applied to the GitHub data). As with the ``SS-*'' examples, these hyperparameters were
selected at random.  
% Lustosa~\cite{lustosa2024learning}. reports that SNEAK performed well on the
% ``*-0'' data sets, but  poorly on the ``*-1'' data sets.

The next two groups in Table~\ref{data} come from the software process modeling literature~\cite{Me07,menzies2009avoid,green2009understanding,port2008using}:
\begin{itemize}
\item The data sets ``pom*'' show data from a model of agile development proposed by Turner and Boehm~\cite{boehm2004balancing}.
In agile projects, developers adjust what to do next   based on their current knowledge of a project. 
POM3 models requirements as a tree of dependencies that rises out of pool of water. At any time, developers can only see the dependencies
above the water. Hence they can be surprised by unexpected dependencies that emerge at a later time. 
POM3 reports the completion rates, idle times, and development effort seen when teams try  navigate this space of changing tasks.  
    \item 
 The data sets ``china'' to ``osp2'' show data in
the format of the USC Cocomo models that predict for development effort, defects, risk in waterfall-style software projects~\cite{Me07}.
``china'' and ``nasa93dem'' contain data from real projects while the others where generated via randomly selected inputs to the USC models.
\end{itemize}
Finally, the last two data sets (``car design'' and ``wine quality'') are included for debugging purposes. All the other data sets
require some specialized SE domain knowledge. These last two data sets lets SE newcomers apply their their general knowledge while implementing
and fixing
these algorithms.

\subsubsection{Experimental Rig}
At the start of each run, the ordering
of the examples in each data set is randomized (to undo any effects from prior runs). These rows are then processed by 
\verb+lite.py+, SWAY, SNEAK, and random(N):
\begin{itemize}
    \item
 \verb+lite.py+'s labelling is controlled by a {\em budget} parameter. From
 Task~\#1, we vary {\em budget}  from 10 to 80. 
\item Random(N) constrains it search using the same {\em budget} parameter as \verb+lite.py+.
\item SWAY and SNEAK set their labelling budgets via the topology of the data.
\end{itemize} 
For each data set, we make 20 runs. Each run returns one {\em d2h} score for the best example found by that method
in that run. 

\subsubsection{Statistical Analysis}

At the end of a simulations study, we will have 20 results per method per {\em budget}. All these are sorted by their mean {\em d2h} then ranked via a Scott-Knot procedure. Scott-Knott recursively partitions the list of candidates ($c$) into two sub-lists ($c_1$ and $c_2$) which the expected mean value before and after the division should be maximized~\cite{emblem, xia2018hyperparameter, 9463120}:
\begin{equation}\label{sk}
    E(\Delta) = \frac{\textrm{len}(c_1) *|\overline{c_1} \overline{c}| + \textrm{len}(c_2) *|\overline{c_2}\overline{c}|}{\textrm{len}(c)}
\end{equation}

\begin{table}[!t]
{\scriptsize
\begin{center} 
\begin{tabular}{c|crrrl}
 statistical & treatment  & required & \multicolumn{3}{c}{{\em d2h} = distance to heaven percentiles in 20 repeats} \\\cline{4-6}   
  rank &  & \#labels &  50th & (75-25)th& (``{\tt o}'' is the 50th percentile median)\\\hline
\rowcolor{gray!20}  &              random&60&  0.07&  0.05& {\verb+ o--+}                                     \\
\rowcolor{gray!20}   &                 certain&30&  0.08&  0.13& {\verb+ o------+}                                 \\
\rowcolor{gray!20}   &              random&80&  0.08&  0.08& {\verb+ o---+}                                    \\
\rowcolor{gray!20}   &                 certain&20&  0.10&  0.13& {\verb+ -o-----+}                                \\
\rowcolor{gray!20}   &              random&50&  0.10&  0.12& {\verb+ -o-----+}                                 \\
 \rowcolor{gray!20} 0&                 certain&40&  0.12&  0.16& {\verb+ ---o-----+}                               \\
\rowcolor{gray!20}   &              random&70&  0.12&  0.09& {\verb+ ---o-+}                                    \\
 \rowcolor{gray!20}  &                 certain&80&  0.12&  0.12& {\verb+ ---o---+}                                \\
 \rowcolor{gray!20}  &              uncertain&70&  0.15&  0.13& {\verb+ ----o---+}                                \\
 \rowcolor{gray!20}  &                 certain&50&  0.16&  0.17& {\verb+ -----o----+}                              \\
 \rowcolor{gray!20}  &              random&40&  0.18&  0.09& {\verb+    ---o-+}                                \\ 
   &                 certain&60&  0.19&  0.12& {\verb+    ----o--+}                              \\
    &              uncertain&80&  0.19&  0.12& {\verb+    ----o--+}                              \\
   &              uncertain&30&  0.20&  0.14& {\verb+ ------o--+}                               \\
    &              uncertain&50&  0.20&  0.09& {\verb+     ---o-+}                               \\
    &              random&30&  0.20&  0.08& {\verb+      --o--+}                               \\
    1&              uncertain&60&  0.20&  0.10& {\verb+     ---o--+}                               \\
    &              uncertain&40&  0.21&  0.15& {\verb+ --------o+}                                \\
   &              uncertain&10&  0.21&  0.09& {\verb+       --o--+}                               \\
    &              uncertain&20&  0.21&  0.12& {\verb+    -----o-+}                              \\
    &              random&20&  0.22&  0.12& {\verb+     ----o--+}                             \\
  &                 certain&70&  0.22&  0.07& {\verb+       --o-+}                              \\ 
\rowcolor{gray!20}  2&             SNEAK&22&  0.22&  0.04& {\verb+        -o-+}                               \\ 
    3&               sway&9&  0.24&  0.15& {\verb+       ---o-----+}                          \\ 
\rowcolor{gray!20}   &                 certain&10&  0.26&  0.11& {\verb+          -o----+}                        \\
\rowcolor{gray!20}  4&              random&10&  0.28&  0.09& {\verb+          ---o- +}                         \\
 \rowcolor{gray!20}  &                sway&5&  0.28&  0.13& {\verb+            -o-----+}                     \\ 
   5&             baseline&500&  0.51&  0.20& {\verb+                    ------o-----+}         \\
\end{tabular}
\end{center}}
\caption{{\em d2h} results for pom3a.  Results are 
sorted by the median {\em d2h} (distance to heaven) as calculated by Equation~\ref{d2h}.   Shading denote statistically indistinguishable results
(as displayed in column one and computed by the Scott-Knott calculation of \S\ref{sk}).  Results from  20 repeats of  studying  ``pom3a'' (this data set has 500 rows and 12 attributes).   Horizontal box plots on the right-hand-side shows median and IQR (75th=25th percentile).
The {\em uncertain} and {\em certain} policies are  defined by Equations~\ref{uncertain} and \ref{certain}. }\label{pom3a1}
\end{table}
\begin{table}[!t]
 {\scriptsize
\begin{center} 
\begin{tabular}{ccrrrl}
 statistical &   & required & \multicolumn{2}{c}{\em d2h} \\\cline{4-5}   
  rank & treatment & \#labels &  50th & (75-25)th \\\hline

  \rowcolor{gray!20}   &                 certain&20&  0.10&  0.13   & \cellcolor{white} $\leftarrow$ best certain                       \\
\rowcolor{gray!20}   &                 certain&30&  0.08&  0.13                              \\
 \rowcolor{gray!20}  &                 certain&40&  0.12&  0.16                            \\
  \rowcolor{gray!20}  &              random&40&  0.18&  0.09     &  \cellcolor{white}$\leftarrow$ best  random                         \\ 
\rowcolor{gray!20} 0 &                 certain&50&  0.16&  0.17                              \\

\rowcolor{gray!20}  &              uncertain&70&  0.15&  0.13    &  \cellcolor{white}$\leftarrow$ best  uncertain                           \\
\rowcolor{gray!20}   &              random&70&  0.12&  0.0                                    \\

  &              uncertain&10&  0.21&  0.09                                \\
   
   1  &              uncertain&40&  0.21&  0.15                                 \\
   &              uncertain&50&  0.20&  0.09                               \\
     
&              uncertain&80&  0.19&  0.12                               \\
 
\rowcolor{gray!20}  2&             SNEAK&22&  0.22&  0.04 & \cellcolor{white}$\leftarrow$ best  SNEAK                            \\ 
    3&               sway&9&  0.24&  0.15     &  \cellcolor{white}  $\leftarrow$ best  SWAY                    \\ 
 \rowcolor{gray!20}  &                sway&5&  0.28&  0.13                     \\ 
 \rowcolor{gray!20}  4&                 certain&10&  0.26&  0.11                         \\
\rowcolor{gray!20}   &              random&10&  0.28&  0.09                          \\

   5&             baseline&500&  0.51&  0.20       \\
\end{tabular}
\end{center}}
\caption{Same data and  format as Table~\ref{pom3a1} but, here, the rows sorted ascending {\em first}  on statistical rank and {\em second} on required number of labels. Less informative rows have been deleted.}\label{pom3a2}
\end{table}

Hypothesis test is then applied to check if two sub-lists differ significantly by using the Cliff's delta procedure. The delta value is \[Delta = (\#(x>y) - \#(x<y))/(\textrm{len}(c_1)*\textrm{len}(c_2)) for \forall x \in c_1 and \forall y \in c_2\]To explain that, Cliff's delta estimates the probability that a value in the sub-list $c_1$ is greater than a value in the sub-list $c_2$, minus the reverse probability~\cite{macbeth2011cliff}. Two sub-lists differ significantly if the delta is not a ``small'' effect ($Delta >= 0.147$)~\cite{hess2004robust}.

The reason we choose Scott-Knott because (a) it is fully non-parametric and (b) it reduces the number of potential errors in the statistical analysis since it only requires at most $O(log2(N))$ statistical tests for the $O(N^2)$ analysis.
Other researchers also advocate for the use of this test~\cite{gates78}   since it   overcomes a common limitation of alternative multiple-comparisons statistical tests (e.g., the Friedman test~\cite{friedman1937use}) where   treatments are assigned to multiple groups (making it hard for an experimenter to   distinguish the real groups  where the means should belong~\cite{Carmer1985PairwiseMC}.

 Table~\ref{pom3a1} illustrates the use of Scott-Knot on the 20 results generated from the  ``pom3a'' dataset. In that table, shaded regions are used
 to highlight the Scott-Knott partitions. 
 \begin{itemize} 
 \item Within each partition, the methods are statistically indistinguishable.
 \item 
The untreated data is shown as the last row: note that  most examples in the ``pom3a'' data set have a distance to heaven of 0.51. 
\item 
The best ranked results in Table~\ref{pom3a1}  are labelled ``0'' and are shown at the top of the table. This group has a median {\em d2h} of 0.12 and the results of the methods in this group are statistically indistinguishable  from the best result (of 0.07). 
\end{itemize}

Table~\ref{pom3a2} resorts  the rows of Table~\ref{pom3a1}. In this second figure, 
 the rows are sorted ascending {\em first}  on statistical rank and {\em second} on required number of labels.
 With that sort, 
the ``best  result'' for each method can be read from the top-most mention of that method (i.e it is the method where the best ranked results are achieved after
reviewing the fewest labels).

\subsection{Results}\label{results}

In this section, we use the methods described about  to answer our research questions.

\subsubsection{ {\bf RQ1:}   What are the least number of labels we should use?}
To compute this, we ask:
\begin{itemize}
    \item  What are the least number of labels needed to find examples which have a  lower distance to heaven...
    \item ...and which are 
statistically distinguishable  from the baseline? 
\end{itemize}
If the baseline is in partition $N$ (e,g, $N=5$  in Table~\ref{pom3a2}) then the smallest size seen in partition $N-1$
is the fewest labels needed to achieve a statistically  distinguishable reduction in {\em d2h}:
\begin{itemize}
    \item   In the case of
Table~\ref{pom3a2}, that number is 5 (see   ``sway''   in partition $N=4$).
\item For all 31 data sets, that number has the value $\mu=12, \sigma=2.4$
\end{itemize}
Hence we say that 12 labels (on average) can be enough to find an effect that changes some baseline
phenomenon by a  statistically distinguishable amount. 
That said,  we do not recommend stopping at 12 since, with just a few more labels, we can achieve much larger reductions 
in {\em d2h}.

\begin{table}[!b]
\centerline{
 \begin{tabular}{r|rrrrr}
\multicolumn{1}{c}{~}&&\multicolumn{2}{c}{{\tt lite.py}}\\\cline{3-4}
\multicolumn{1}{c}{} & random& certain              & uncertain              & sway                  & sneak  \\
\multicolumn{1}{c}{statistical rank}                      & $\mu ~ (\sigma)$ & $\mu ~ (\sigma)$ &  $\mu ~ (\sigma)$ & $\mu ~ (\sigma)$ & $\mu ~ (\sigma)$  \\\hline  
best $\rightarrow$ 0  & \cellcolor{CustomDarkRed!96} \textcolor{white}{46 (26)} & \cellcolor{CustomDarkRed!65} \textcolor{white}{30 (25)}& \cellcolor{CustomDarkRed!41} 32 (22)& \cellcolor{CustomDarkRed!9}  7 (5)& \cellcolor{CustomDarkRed!12} 81 (43)    \\
1                     &   & \cellcolor{CustomDarkRed!6 } 50 (19)& \cellcolor{CustomDarkRed!12}   30 (13)& \cellcolor{CustomDarkRed!6} 7 (5)    & \cellcolor{CustomDarkRed!16} 83 (42)\\
2                     &   & \cellcolor{CustomDarkRed!12} 40 (19)& \cellcolor{CustomDarkRed!29} 27 (16)& \cellcolor{CustomDarkRed!16}  6 (5)   & \cellcolor{CustomDarkRed!16} 51 (36)\\
3                     &  & \cellcolor{CustomDarkRed!6} 20 (19)& \cellcolor{CustomDarkRed!6} 45 (54)& \cellcolor{CustomDarkRed!32}  6 (5)  & \cellcolor{CustomDarkRed!29} 24 (14)\\
4                    & & \cellcolor{CustomDarkRed!3} 50 (5) & \cellcolor{CustomDarkRed!16} 55 (12)& \cellcolor{CustomDarkRed!19}  6 (6)& \cellcolor{CustomDarkRed!12} 7 (8)\\
5                    & &                            &                             & \cellcolor{CustomDarkRed!6}  4 (5)& \cellcolor{CustomDarkRed!9}   63 (38)\\
6                   &  &                             &                             & \cellcolor{CustomDarkRed!3}    4 (5)&  \\
worst $\rightarrow$ 7    &                         &                          & \cellcolor{CustomDarkRed!3}    5 (5)&  
~\\\multicolumn{1}{c}{~}\\
{\bf LEGEND:} &
\cellcolor{CustomDarkRed!95} \textcolor{white}{95}  & \cellcolor{CustomDarkRed!65} \textcolor{white}{65\%}  & \cellcolor{CustomDarkRed!50} \textcolor{white}{50\%}  & \cellcolor{CustomDarkRed!25} 25\%  & 
\cellcolor{CustomDarkRed!5} 5\%    

\end{tabular}}
\caption{Results from 20 repeats over 31 data sets. 
\underline{\bf Numbers} denote mean (and standard deviation) of number of labels
required to achieve a ``best result'' at the left-hand-side statistical ranks.Here {\em smaller} numbers are {\em better} (and ``best result'' is defined as per Table~\ref{pom3a2}).
\underline{\bf Cell shading} denote how often (in 31 data sets) did a method
achieve results at that rank. Cells with 
{\em darker} shades occur {\em more often}.
For example in 67\% of the results,
{\em certain} sampling with {\tt lite,py} reached the best rank ``0''
after using just $\mu=30$ labels. 
}\label{percent}
\end{table}

\subsubsection{{\bf RQ2:} {\em What are the most number of labels we should use?}}

Much  current work in analytics assumes   ``the more {\eg}s the 
better''~\cite{mcintosh2017fix,rahman2013sample,amasaki2020cross}.  Perhaps this is true in general   but in the specific case of labelling examples for software review, our results offer an alternate perspective.

Equation~\ref{hamlet} promised that  
49 labels could suffice for finding examples that are
statistically indistinguishable from the best result.
This theoretical result   is remarkably close to what we see in  practice:
\begin{itemize}
\item 
Table~\ref{percent}  shows how often different methods achieve ``best results'' in our 31 data sets   (where ``best result'' is defined via  Table~\ref{pom3a2}). 
\item 
Top left   of  Table~\ref{percent},   we see that  random sampling does very well indeed.
Specifically, 
in 80\% of our 31 data sets, to find examples
that are indistinguishable from   best. 
it is enough to look at 
46 labels (on average).
\end{itemize}
Note that  we can often do better than ``look at 46 labels''.  The {\em certain} results
of \verb+lite.py+ in Table~\ref{percent}  show that, 67\% of the time,
that method can  find examples
that are indistinguishable from   best by looking at just 30 labels (on average).
 
We hence conclude that it is not always true that more labels are always useful.
After  2-4 dozen labels, there are very sharply diminishing returns
in looking at further labels.
 
{\bf RQ3}: {\em  What are the merits of purely random labelling?}

As mentioned above, random labelling is very simple to implement but does not deliver a model
that can help to reduce the effort associated with future examples.  However,
Table~\ref{percent} shows that
two-thirds the time, methods like \verb+lite.py+(certain) do just as well as random.
Further, random(N) has no way to incorporate SME knowledge (if that knowledge is available). 
 On the other hand, methods like
\verb+lite.py+   could check and change its conclusions
with a human at each step in its sampling loop.

\subsection{ {\bf RQ4:} {\em For non-random labelling, what policy should be used ?}} 

 The active learning literature  (e.g.~\cite{muller22}) usually recommends the uncertainty sampling of  Equation~\ref{uncertain}. Does that recommendation hold here?

 Looking across the columns of Table~\ref{percent},  we see that \verb+lite.py+ with certain sampling has more red cells, higher in the table, that the other methods.
 This is to say that \verb+lite.py+  finds higher statistically ranked results much more often than uncertain sampling, SWAY, or SNEAK. Hence, for software review:
 \begin{itemize}
     \item  We recommend certainty sampling rather than  uncertainty sampling;
\item 
 We deprecate the greedy search of SWAY and the   non-local pruning technique explored in in SNEAK.
\end{itemize}

 \section{Discussion}\label{discuss}

\subsection{Threats to Validity}
As with any empirical study, different biases can threaten the final results. Therefore, any conclusions born from this work must be considered with the following biases in mind.

\noindent \textbf{Parameter Bias} - \verb+lite.py+ contains multiple different other hyperparameters that can be modified to better fit specific datasets. While for the purposes of this study we demonstrate that the current configuration of this tool is capable of producing outstanding results in the 31 datasets studied here, maybe other parameter settings may be necessary for different tasks and datasets (e.g.: one could attempt to define an early stopping rule for this tool). Another point to be considered here are the two policies ``certain'' and ``uncertain'' present in \verb+lite.py+, there may exist different policies that can yield better performance. These tasks however are left for future work. 

\noindent \textbf{Sampling Bias} - Any empirical study using datasets is threatened by sampling bias, which basically is to say, what works here may not work everywhere. In this study we have applied \verb+lite.py+ to 31 datasets focusing wildly different tasks, but still these results may not hold outside of SE data and on different sub tasks within SE. Other tasks and categories of data need to be evaluated in the future.

\noindent \textbf{Algorithm Bias} - We baselined \verb+lite.py+ against 2 other state of the art algorithms in SWAY and SNEAK as well as a random selection naive baseline. However there might exist different categories of algorithms outside of our purview, that can provide a new and welcome challenge to \verb+lite.py+. To the best of our knowledge the algorithms compared here were the only ones to solve the issue of software review, hence the choice of baselines.
 
\subsection{Implications for Science}
The {\bf RQ1} results of this paper offer some insight into the nature of scientific research. 
How often do research findings emerge from aimlessly tackling a problem until, after examining twelve instances, scholars identify an effect that significantly alters a phenomenon?
How many ``refutation papers'' are just ``we looked at another 12 examples and found a different effect''?  And how  many such studies would be invalidated by a more meticulous choice of examples to investigate?

\subsection{Frequently Asked Questions}
As to other matters, when discussing this paper with colleagues, they are often startled by our results.
 Hence we are often asked ``why does  this approach work so well, with so little data?'' and ``what did prior work get so wrong?''.
For these questions, we have several answers:

\begin{itemize}
    \item 
    Here we are solving optimization problems. 
{\em Optimization} is different to other problems (like 
{\em classification, regression}, or {\em generation}) in that other problems need enough data to apply labels (or weights) across many examples.  Optimization, on the other hand, is the art of throwing things away to find a small number of examples that are better than the rest. If each labelling is considered an experiment that throws away (say) half the   space
of examples, then   $N$ labels could send us to a specific $1/(2^{N})$ part of the data. 
Assuming
$N=30$ labels, this would lead us to a tiny one-billionth  region of  the data where, indeed,
we might find good solutions.

\item
We have read widely in this field of active learning and semi-supervised learning. Much of that work assumes discrete target classes. The challenge with incremental learning for discrete classes is that until one of the target classes appears, the learner has nothing to go on. But here, we deal with continuous classes (see the {\em distance to heaven} metric of Equation~\ref{d2h}). We conjecture that incremental learning is easier in continuous space since any label might be able
to nudge the reasoning ``down the slope'' towards
a better solution.
 \item
We make no guarantees that our solutions are optimal.
We are not finding the best solutions
but, rather,  solutions that are statistically indistinguishable from the best.  As seen in partion ``0'' of
Table~\ref{pom3a1}, there can be many  solutions that  share the same space as ``best''.
Hence, we might not use the methods of this paper  for safety critical applications. 
When the cost of failure is high, there may be enough budget (and time) to label many examples
and the methods of this paper might not be relevant.
\end{itemize}

\section{Conclusion}

In our view,  the culture of software engineering       tend
to   reward   complexity over simplicity. As Edsger Dijkstra warns:
\begin{quote}{\em 
``Simplicity is a great virtue but it requires hard work to achieve it and education to appreciate it. And to make matters worse: complexity sells better.''}
\end{quote}
In this era of readily available executables available in containers, or downloaded as (say) Python/Javascript packages,  it is all too easy to generate a seemingly impressive (and marketable) new solution by mashing up several  prior methods into one complex assembly.   Hence, simplifications studies like this paper are not so common. 

In this simplification study, we have looked into how little do we need to sample software behavior in order to find ways to improve that software. We have argued that it will become an increasingly common concern,
The more we use AI in SE,
the more code will be auto-generated. The more we auto-generate code, the
less time software engineers spend writing and reviewing new code, written by
someone or something else (the internals of which they may not understand). 

Accordingly, this paper has explored a process we call ``software review'' where some panel of subject matter experts spend some time (but not much time) trying to find a way to best use some software. We have characterized such review as a labelling problem with the constraint that we cannot expect many labels. We have explored the labelling problem and shown that  
numerous problems.  that previously, seem to need thousands of millions of labels, can be solved using just a handful
of labels.

\section{Future Work}
We hope that this work is  paper\#1 in a long stream of papers. What might those papers explore? Here are some suggestions:
\begin{itemize}
\item Trials with humans. All the work here was simulations studies to
narrow down the range of possible algorithms. Now that that is done, then
trials with humans would be useful. We warn that such trials have to match some
data/model to humans with expertise in that domain. In practice,
this means that such human trials might have to be much smaller in scale
than the broad ranging inquiry explored here. 
    \item
 Right now, our best methods need around 30 labels to work. 
Can we reduce that? 
\begin{itemize}
\item As we incrementally label examples and incrementally build our classifiers inside \verb+lite.py+, can we recognize early stopping rules?
    \item 
  Also,   this paper has explored two sample policies
(uncertainty sample from Equation~\ref{uncertain} and certainty sample from Equation~\ref{certain}). There are so many other ways to guide the sampling that  are worth exploring

\end{itemize}
\item
Here we assume that all the labels are accurate. What if our labelling oracle is X\% inaccurate? Are their different labelling methods that are more/less susceptible to such noise?
\item
Further, here we assume that our oracles offer one set of labels per example. That might be true when working with one person, but what about teams of humans who might have different opinions on whether or not one example is better than another?
\item
Lastly, our studies here have explored the ``cold start'' problem where we have no background model. But at the end of all the studies reported here, that is no longer true since
now \verb+lite.py+ has models that can predicted for ``best'' or ``rest''. If we came back to this data in a second review session, how can we best lever the model from session\#1? Can we check if the old knowledge is still relevant? Can we combine that old knowledge with new labels? When is it prudent {\em not} to label new data? 
  \end{itemize}

\section*{Acknowledgement}
This work was funded in part by an NSF grant \#1908762.

\section*{Conflicts of Interest Statement}
The authors declare that they have no conflict of interest.

\bibliographystyle{spmpsci}      % mathematics and physical sciences

\bibliography{main} 

\end{document}